\documentclass[lettersize,journal]{IEEEtran}
\usepackage{amsmath,amsfonts}
\usepackage[english]{babel}
\usepackage{amsthm}
\usepackage{algorithmic}
\usepackage{algorithm2e}
\usepackage{fancyhdr}
\usepackage{array}
\usepackage[caption=false,font=normalsize,labelfont=sf,textfont=sf]{subfig}
\usepackage{textcomp}
\usepackage{stfloats}
\usepackage{verbatim}
\usepackage{graphicx}
\usepackage{balance}
\usepackage{tikz}
\usepackage{booktabs}
\usepackage{multicol}
\usepackage{multirow}
\usepackage{cite}
\usepackage{url}
\usepackage{hyperref}
\hypersetup{
    colorlinks=true,
    linkcolor=blue,
    citecolor=blue}
\usepackage{xspace}

\usepackage{marginnote}

\usepackage{xcolor}
\theoremstyle{definition}
\newtheorem{definition}{Definition}

\usepackage{xcolor}
\usepackage{colortbl}  
\usepackage{array}
\makeatletter
\DeclareRobustCommand\onedot{\futurelet\@let@token\@onedot}
\def\@onedot{\ifx\@let@token.\else.\null\fi\xspace}

\def\etal{\emph{et al}\onedot}
\makeatother

\begin{document}
\title{WARBERT: A Hierarchical BERT-based Model for Web API Recommendation}

\author{
Zishuo~Xu,
Yuhong~Gu,
Dezhong~Yao,~\IEEEmembership{Member, IEEE}

\thanks{Zishuo Xu is with the School of Software Engineering and Dezhong Yao is with the School of Computer Science and Technology, Huazhong University of Science and Technology, Wuhan 430074, China (E-mail: ZishuoXuHUST@163.com, dyao@hust.edu.cn).}
\thanks{Yuhong Gu is with the SmartX, Beijing 100086, China (E-mail: yuhong.gu@smartx.com).}
\thanks{\protect (Corresponding author: Dezhong Yao) }
}

\maketitle
\begin{abstract}
  With the rise of Web 2.0 and microservices, the increasing availability of Web APIs has intensified the need for effective recommendation systems. Existing approaches are generally categorized into two methods: recommendation-type methods, which classify APIs using labels, and match-type methods, which retrieve APIs through matching with mashups. However, three significant challenges remain: 1) semantic ambiguities in comparing API and mashup descriptions, 2) a lack of progressive
semantic refinement between mashup requirements and individual API
descriptions, and 3) computational inefficiency of exhaustive
mashup-API comparisons in large-scale repositories.
 To tackle these challenges, we propose WARBERT, a hierarchical model based on BERT for Web API recommendation. WARBERT utilizes dual-component feature fusion and attention mechanisms to create accurate semantic representations. It consists of WARBERT(R) for initial candidate filtering using recommendation methods, and WARBERT(M), which focuses on refined similarity matching. The final likelihood of an API-mashup pairing combines predictions from both components, with WARBERT(R) further enhanced by an auxiliary task of predicting mashup categories.
 Experiments conducted on the ProgrammableWeb dataset demonstrate WARBERT outperforms existing baselines, achieving notable improvements in both accuracy and efficiency.

\end{abstract}


\section{Introduction}\label{sec:introduction}
\IEEEPARstart  {W}{ith} the rapid development of Web 2.0 and microservice architectures, Web APIs have become fundamental building blocks for modern application development~\cite{zhao2020distributed}.
Since 2022, public repositories such as ProgrammableWeb report over 24,000 APIs offering diverse functionalities~\cite{GongZC0B0YQ22}. These resources have given rise to a novel application paradigm known as \textit{Mashup}~\cite{yu2008understanding}, which allows developers to integrate existing Web APIs to meet complex requirements without building applications from scratch~\cite{10542468, ding2023joint}. 
While mashups significantly reduce development cost and time-to-market, they introduce a critical challenge: how to accurately and efficiently identify APIs that best match a developer’s functional requirements from a large and heterogeneous repository.

To address this challenge, Web API recommendation has been extensively studied within the service computing community~\cite{wang2023functional, WangXLPWDY24}. Given a natural-language description of a mashup’s requirements, a recommendation system aims to identify and rank APIs whose functionalities are most relevant~\cite{sang2023mashup,wang2025r2api}.

\begin{figure}[t]
\centering
\includegraphics[width=\linewidth]{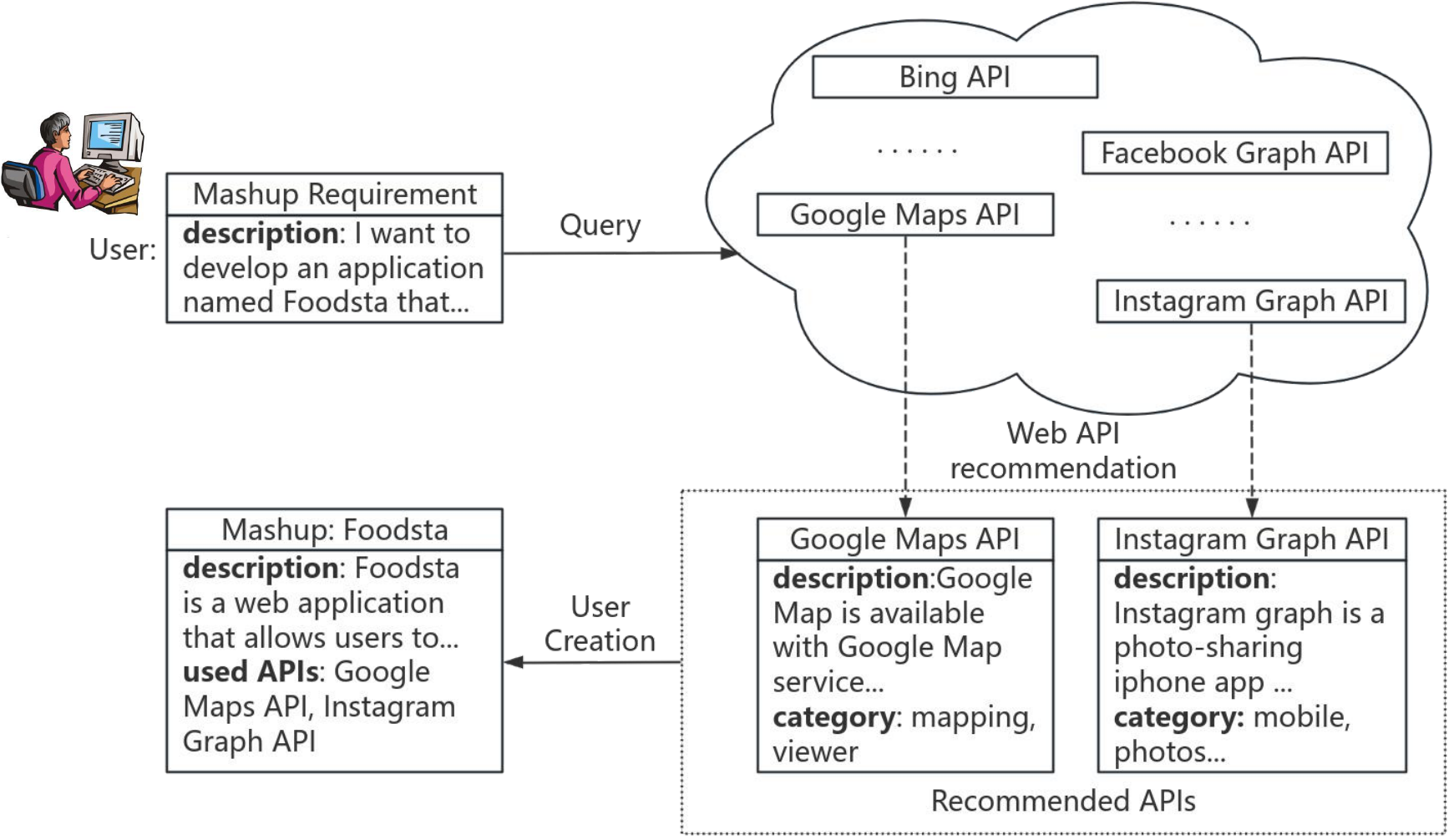}
\caption{An example of a Web API recommendation.}
\label{fig:ar}
\end{figure}

Figure~\ref{fig:ar} illustrates this process with the ``Foodsta'' mashup: the recommendation system facilitates decision-making by assessing whether API functionalities align with mashup needs, considering requirement descriptions, API descriptions, and categories. 
Existing Web API recommendation approaches can be broadly categorized into \textit{recommendation-type methods} and \textit{match-type methods}, depending on how mashup requirements and APIs are modeled and compared. 

\textit{Recommendation-type methods} formulate API recommendation as a multi-label classification problem~\cite{kang2024ks}, where each API corresponds to a label and the model predicts the likelihood that a mashup will invoke each API. Representative approaches include CNN-based models such as MTFM~\cite{wu2021mashup} and BERT-based classifiers such as BERT-CM~\cite{zhang2024cooperative}. These methods are effective at capturing global relevance patterns between mashups and APIs. However, their modeling paradigm inherently limits semantic expressiveness. Since mashup descriptions are encoded independently and mapped directly to a fixed API label space, recommendation-type methods do not explicitly perform pairwise semantic comparison between mashups and individual APIs. Moreover, their classifier structures are tightly coupled to the size of the API repository, which reduces flexibility when APIs are frequently added or removed. As a result, these methods struggle to support fine-grained semantic analysis and dynamic service ecosystems.

\textit{Match-type methods} treat Web API recommendation as a retrieval problem~\cite{kang2024ks}, explicitly computing semantic similarity between mashup descriptions and API descriptions. Recent approaches leverage deep neural networks, attention mechanisms, and pre-trained language models, such as ServiceBERT~\cite{wang2021servicebert}, to enhance semantic representation quality. Although match-type methods enable more detailed semantic alignment, they incur significant computational overhead, as each mashup must be matched against all candidate APIs. In addition, many existing match-type models encode mashup and API descriptions independently and apply similarity measures afterward, which limits their ability to resolve semantic discrepancies arising from different description styles used by service consumers and API providers.

The above analysis reveals that the limitations of existing approaches are structural consequences of their modeling paradigms, rather than implementation deficiencies. In particular, three challenges remain unresolved. 1) \textbf{Semantic Ambiguity Between Mashups and APIs}: Mashup developers and API providers often describe similar functionalities using different terminology, while the same term may convey distinct meanings in different contexts. This makes it difficult to embed mashup and API descriptions into a unified semantic space that reliably captures functional relevance. 2) \textbf{Lack of Progressive Semantic Refinement:} Existing approaches directly predict API relevance without explicitly modeling semantic interactions between mashup requirements and API descriptions. This limits their ability to justify or refine recommendations at the individual API level. 3) \textbf {Inefficiency in Large-Scale API Repositories:} Exhaustive mashup–API comparisons are computationally expensive and memory-intensive, especially as the number of APIs grows.

To tackle the above challenges, we propose WARBERT\footnote{We have open source code at \url{https://github.com/ZsXu-enrico/WARBERT}}, a hierarchical \underline{BERT}-based model for \underline{W}eb \underline{A}PI \underline{R}ecommendation. WARBERT consists of two components: WARBERT(R) and  WARBERT(M). WARBERT(R) formulates API recommendation as a multi-label classification task and serves as an efficient filter that narrows the candidate API set. WARBERT(M) performs fine-grained semantic matching between a mashup and the filtered APIs by jointly encoding their descriptions and explicitly modeling cross-text interactions. The final recommendation score is obtained by fusing the outputs of both components within a unified hierarchical pipeline.

To enhance semantic representation while maintaining efficiency, WARBERT adopts a lightweight BERT (BERT-Tiny~\cite{bhargava2021generalization,DBLP:journals/corr/abs-1908-08962}), and introduces a dual-component feature fusion strategy that combines complementary sentence-level representations. In addition, an attention-based comparison mechanism is employed in WARBERT(M) to highlight discriminative terms in API descriptions conditioned on mashup requirements. WARBERT(R) is further augmented with an auxiliary mashup category prediction task, which provides additional semantic supervision and improves representation quality. Extensive experiments on the ProgrammableWeb dataset demonstrate that WARBERT consistently outperforms state-of-the-art baselines in terms of accuracy and ranking quality, while maintaining practical inference efficiency. Ablation studies further validate the effectiveness of the hierarchical architecture, feature fusion strategy, and attention-based comparison mechanism.
The main contributions of this work are summarized as follows:
\
\begin{itemize}
    \item We propose WARBERT, a novel hierarchical framework that systematically integrates recommendation-type filtering and match-type semantic matching for Web API recommendation.

    \item We design a dual-component feature fusion strategy and an attention-based comparison mechanism to improve semantic alignment between mashup requirements and API functionalities.
    

    \item We introduce an efficient filtering–matching pipeline based on lightweight BERT, achieving a favorable balance between effectiveness and computational cost.
    

    \item We conduct extensive experiments and ablation studies demonstrating that WARBERT achieves superior performance over existing methods in both API recommendation and mashup category prediction tasks.


\end{itemize}


\section{Related Work}
\label{sec:relwork}
Web API recommendation has been extensively studied in the service computing community. Existing methods can be broadly categorized into content-based methods, collaborative filtering (CF) methods, factorization machine–based (FM) methods, and hybrid approaches.

\subsection{Content-Based Methods}
Content-based methods recommend APIs by analyzing textual information such as mashup descriptions and API documentation~\cite{CaoPZQTKL23}. These approaches can be further divided into recommendation-type and match-type methods.

Recommendation-type methods formulate API recommendation as a classification problem, where each API is treated as a label. Early approaches relied on keyword matching or topic modeling, while recent studies leverage deep neural networks to extract semantic features from textual descriptions\cite{10.1007/s10115-024-02061-2}. With the emergence of pre-trained language models, BERT-based encoders have been employed to obtain contextualized representations of mashup descriptions, followed by classifier layers to predict API relevance, such as BERT-CM~\cite{zhang2024cooperative}. Although these methods are efficient at inference time, they typically produce relevance scores without explicitly modeling fine-grained interactions between mashups and individual APIs, which limits their expressiveness.

Match-type methods instead model API recommendation as a retrieval or matching problem. These approaches compute similarity scores between mashup descriptions and API descriptions using neural networks, attention mechanisms, or contrastive learning. Zhong \etal~\cite{ZhongF0Z18} reconstruct service profiles from mashup descriptions and ranked candidates by conditional probability with dominant-word boosting. Chen \etal~\cite{9780620} introduce a keyword-driven service recommendation using deep reinforced Steiner tree search. FC-LSTM~\cite{shi2019functional} implements an attention mechanism based on LSTM which enables the model to give higher weight to more important descriptions. ServiceBERT~\cite{wang2021servicebert} employs pre-training and contrastive learning to extract the semantic embeddings of mashup descriptions and API descriptions via BERT. DLOAR~\cite{Wang2023DeepLO} builds mashup topic models via RoBERTa and clustering, extracts keywords with TextRank, and ranks open APIs through three-level neural similarity.
While match-type methods enable detailed semantic comparison and are flexible to changes in the API repository, they often incur high computational overhead due to exhaustive pairwise matching and lack efficient candidate filtering strategies.

\subsection{Collaborative Filtering Based Methods}
Collaborative filtering (CF) methods exploit historical mashup–API invocation records to capture usage patterns and implicit relationships~\cite{4052914}.
Cao \etal~\cite{buqing2014cscf} combine content similarity and collaborative filtering for Web API recommendation. Chen \etal~\cite{Chen2014QoSAwareWS} propose a collaborative filtering approach that integrates user-based CF with item-based CF to identify suitable APIs by evaluating QoS information. A geographic-aware CF strategy was proposed by Botangen \etal~\cite{BotangenYSHY20}. Matrix factorization paired with network maps is utilized by Tang \etal~\cite{7378981} to predict web service quality. Yao \etal~\cite{YaoWSBH21} introduce a probabilistic matrix factorization (PMF) approach for mashup recommendation, focusing on the effects of explicit similarities and implicit API correlations on API co-invocation. LNGCF \cite{xiang2023interactive} captures high-order interaction dependencies through linear propagation on user-API graphs and dynamically aggregates multi-layer embeddings to enhance performance. However, CF-based methods are highly sensitive to data sparsity and cold-start problems, and they typically struggle to leverage rich textual semantics from mashup and API descriptions.

\subsection{Factorization Machine Based Methods}
Factorization machines (FMs) are effective at modeling high-order feature interactions in sparse settings. In the context of Web API recommendation, FM-based approaches integrate semantic features extracted from text with interaction features derived from mashup–API networks.
Xie \etal~\cite{XieCYZL18} use similarity matrices from heterogeneous networks and meta-paths to extract features for Web services and Mashups. 
Anelli \etal~\cite{AnelliNSRT19} improve the interpretability of factorization machines in recommendation systems by initializing latent factors with semantic features from knowledge graphs. Both NAFM~\cite{kang2020nafm} and AMF~\cite{nguyen2021attentional} incorporate attention mechanisms to capture significant parts of the latent features. In addition, the FCN-NAFM model~\cite{kang2024web} further enhances recommendation performance by integrating content information with network structure information of Web services. Cao \etal~\cite{10372383} leverage xDeepFM's compressed Interaction Network to explicitly model high-order feature interactions while integrating GNN for structural modeling. Despite their effectiveness, FM-based models often rely on carefully engineered features and may have limited ability to capture deep semantic alignment between natural-language descriptions.

\subsection{Hybrid Methods}
 
Hybrid methods aim to combine complementary information sources, such as textual content, invocation history, and structural relationships. RWR~\cite{wang2018mashup} computes API relevance via random walk with restart on a heterogeneous knowledge graph of tags, categories, and mashups from a query boundary subgraph with heuristic weighting. Xie \etal~\cite{xie2019personalized} develop a complex Heterogeneous Information Network for tailored Web API recommendations for mashups.
The MTFM model \cite{wu2021mashup} explores both the semantics of mashups and the generation of latent interactions between mashups and APIs.
FSFM \etal~\cite{wang2023functional} extracts embedding vectors and generates requirement semantic vectors for Mashups and APIs, ultimately fuses all representation vectors to produce a list of recommended APIs. 
 Tang \etal~\cite{10.1016/j.infsof.2024.107428} leverage pre-trained heterogeneous information networks to encode semantic and structural relationships. SEHGN~\cite{WangXLPWDY24} combines heterogeneous graph neural networks with semantic embedding techniques to model complex structural dependencies in mashup-API interaction networks and  textual descriptions. LLMAR~\cite{10948476} combines multi-task instruction learning to formulate API recommendation as a generative task. SRCA~\cite{ZOU2025104219} uses prompt-enhanced LLMs to
extract semantic representations 
and constructs a category co-occurrence graph to capture implicit functional
correlations. However, many hybrid methods introduce significant model complexity or computational overhead, and few explicitly address the trade-off between fine-grained semantic matching and scalable inference.

\subsection{Positioning of This Work}
Unlike prior approaches that focus exclusively on either classification-based recommendation or pairwise semantic matching, WARBERT adopts a hierarchical filtering–matching paradigm that integrates both perspectives within a unified framework. By using a recommendation-type model to efficiently prune the candidate space and a match-type model to perform fine-grained semantic comparison, WARBERT achieves a balance between effectiveness and efficiency. Moreover, the use of lightweight BERT, dual-component feature fusion, and attention-based comparison distinguishes WARBERT from existing hybrid solutions that rely on heavier models or structural assumptions.

\section{Problem Definition}
\label{sec:prodef}
In this section, we first formally state the Web API recommendation task and then explain the role of WARBERT in the overall Web API recommendation workflow.

\subsection{The Definition of Web API Recommendation}
 This paper focuses on the Web API recommendation task for mashup applications by providing suitable APIs within a cloud environment. Let a requirement description for mashup $m_i$ provided by a user be denoted as $T^{m_i}$. Let $A = \{{a_1}, \ldots, {a_j}, \ldots, {a_L}\}$ represent a repository of $L$ Web APIs, where each Web API ${a_j}$ is characterized by function description, category and meta-elements. Given $T^{m_i}$ and $A$, the Web API recommendation task aims to generate a ranked subset $S^{m_i} \subseteq A$: $ S^{m_i}=\{a_{\sigma(1)},  \ldots, a_{\sigma(N)}\}$, 
 where $\sigma$ is a permutation function that ranks APIs in $A$ according to their match score with $T^{m_i}$, and $N$ is the number of the selected APIs. Then the user can develop their mashup via APIs in $S^{m_i}$.

\begin{definition}[Requirement description]
$T^{m_i}$ is an unstructured text to represent a
user requirement for mashup creation.
\end{definition}

\begin{definition}[Web API]
$a_j = \langle D^{a_j}, C^{a_j},X  \rangle \in A$ consists of: a textual description $D^{a_j}$ to introduce its functions,
    a set of categories $C^{a_j}$ to index it,
   a set of meta elements $X$, such as quality attributes, input ports, and
    output ports. Each API has at least either an input or an output port \cite{zhang2024cooperative,imran2012systematic}.
\end{definition}

\begin{definition}[Web API Recommendation Task]
Given the requirement description $T^{m_i}$ and the API repository $A$, the Web API recommendation task is to compute a subset $S^{m_i}$ with selected APIs. The objective of the task is to maximize the match score between $T^{m_i}$ and the APIs in $S^{m_i}$: \begin{align}
S^{m_i} = \operatorname*{arg\,max}_{\substack{S \subseteq {A}, \\ |S| = N}} \sum_{a_j \in S} r^{m_i}_j,
\end{align}
where $r^{m_i}_j \in [0,1]$ is the match score between requirement description $T^{m_i}$ and the API $a_j$.
\end{definition}

\begin{definition}[Mashup]
$m_i = \langle D^{m_i}, C^{m_i}, A^{m_i}, X \rangle \in M$ consists of: a textual description $D^{m_i}$ to introduce its functions,
    a set of categories $C^{m_i}$ to indicate its functions,
    a bundle of APIs $A^{m_i}$ to participate in the mashup,
    a set of meta-elements $X$ to prepare for composition \cite{zhang2024cooperative,imran2012systematic}.
\end{definition}

\begin{definition}[Web API Recommendation Model]
Web API recommendation Model is designed to handle the Web API recommendation task. It is defined as a parameterized function which outputs match score vector $R^{m_i}$: $f(\theta, T^{m_i}, A) \rightarrow R^{m_i}=[r^{m_i}_1,\ldots, r^{m_i}_j, \ldots r^{m_i}_{L}]$. By applying the permutation function $\sigma$ on $R^{m_i}$ and selecting the top-$N$ APIs, the subset $S^{m_i}$ is obtained.

\end{definition}

\begin{figure}[t]
\centering
\includegraphics[width=\linewidth]{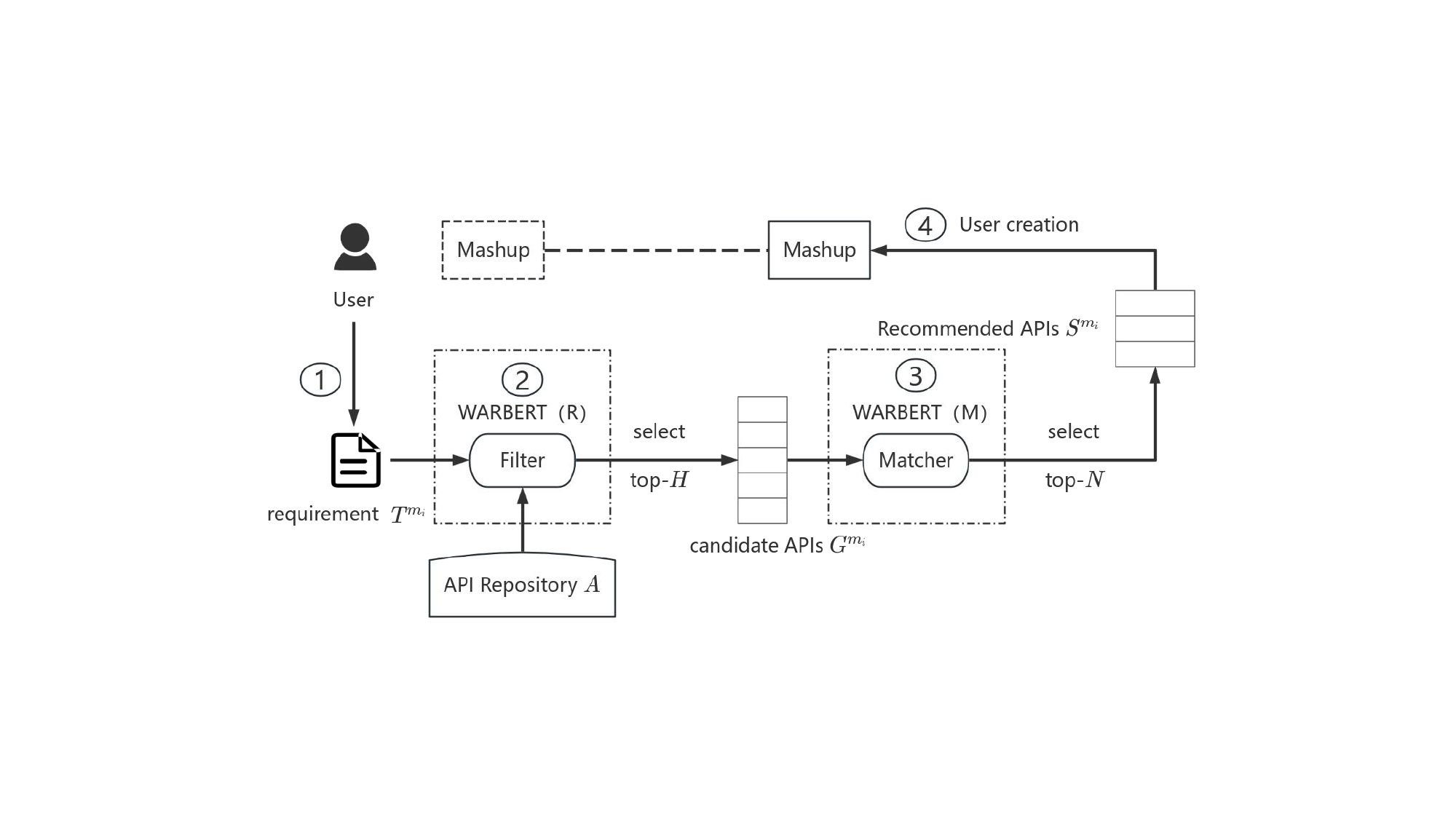}
\caption{The workflow of Web API recommendation.}
\label{fig:wf_m}
\end{figure}

\subsection{The Workflow of Web API Recommendation }
 Figure~\ref{fig:wf_m} illustrates the whole workflow. First, the user proposes a new mashup requirement $T^{m_i}$. Second, the requirement is processed through the filter to select the top-$H$ candidate APIs from the whole API repository. Third, the matcher performs deep matching between the requirement and the candidate APIs $G^{m_i}$, and ranks them to produce the final recommendation results $S^{m_i}$. Finally, the user develops the mashup using the recommended APIs. The filter eliminates APIs that are unlikely to be invoked by the mashup, thereby improving efficiency.

\section{Methodology}
\label{sec:method}
In this section, we first outline the overall architecture of WARBERT and describe the implementation of WARBERT for both components, then analyze its training process. 

\subsection{WARBERT Architecture}

\begin{figure*}[!t]
\centering
\includegraphics[width=1\linewidth]{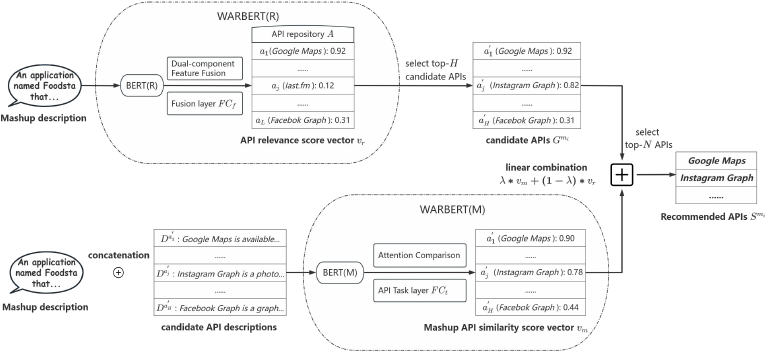}
 \caption{The overview of WARBERT framework with a use case of the mashup ``Foodsta''.} 
\label{fig:frame}
\end{figure*}

\begin{figure}[t]
\centering

\includegraphics[width=\linewidth]{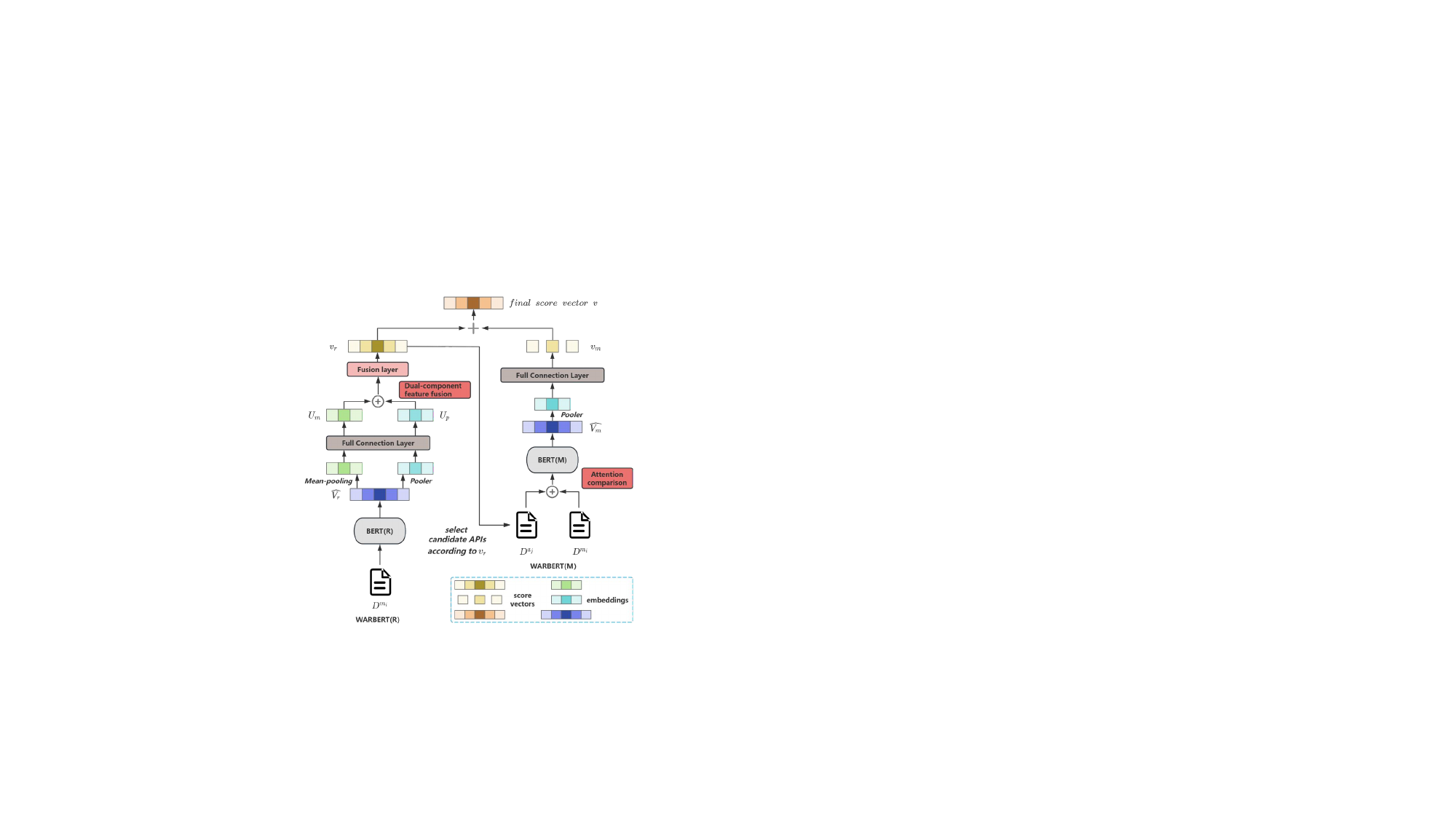}
\caption{The detailed architecture of WARBERT.}
\label{fig:arch}
\end{figure}

Figure~\ref{fig:frame} shows the overall framework of WARBERT, 
Figure~\ref{fig:arch} illustrates the detailed architecture of its two
components and Table~\ref{tab:notation} summarizes the key 
notations used in the model. WARBERT consists of WARBERT(R) and WARBERT(M). WARBERT(R)
treats API recommendation as a multi-label classification problem, where
each API corresponds to a label. It employs dual-component feature fusion
to combine complementary semantic representations for richer encoding.
WARBERT(M) treats API recommendation as a retrieval process, concatenating
mashup and API descriptions as a joint input to enable attention-based
comparison that highlights functionally relevant terms. Both components adopt BERT as the backbone since BERT captures rich semantic and generate contextual embeddings. In the hierarchical architecture, WARBERT(R) functions as a filter to select potential API candidates, and WARBERT(M) then performs fine-grained matching between the mashup and these candidates. The final matching result is a weighted average of the API relevance score vector predicted by WARBERT(R) and the mashup API similarity score vector predicted by WARBERT(M). Ultimately,  we rank and choose the top-$N$ APIs with the highest match scores as the recommended set.

\begin{table}[t]
\centering
\caption{Summary of Notations}
\label{tab:notation}
\small
\begin{tabular}{cl}
\toprule
\textbf{Notation} & \textbf{Description} \\
\midrule
$m_i$ & The $i$-th mashup \\
$a_j$ / $a_{j'}$ & The $j$-th API / $j'$-th candidate API \\
$D^{m_i}$ & Word sequence of mashup $m_i$ description \\
$D^{a_j}$ & Word sequence of API $a_j$ description \\
$L$ & Total number of APIs in the repository \\
$H$ & Number of candidate APIs after filtering \\
$F$ & Dimension of BERT embeddings \\
$\widehat{V_r}$ & Contextual embedding in WARBERT(R) \\
$\widehat{V_m}$ & Joint contextual embedding in WARBERT(M) \\
$U_p$ / $U_m$ & Feature vectors from pooler-output / mean-pooling \\
$v_r$ & API relevance score vector from WARBERT(R) \\
$v_m$ & Similarity score vector from WARBERT(M) \\
$v$ & Final score vector \\
$\lambda$ & Weighting factor for score fusion \\
\bottomrule
\end{tabular}
\end{table}

\subsection{Implementation of WARBERT(R)}
\subsubsection{Computing Contextual Embedding}


Given the mashup description and API description, the word sequences
are obtained through tokenization. The word sequence of the mashup $m_i$ description is denoted as $D^{m_i} =[d_1^{m_i}, ..., d_{n_t}^{m_i}]$, while the word sequence of the API $a_j$ description is denoted as $D^{a_j} =[d_1^{a_j}, ..., d_{n_t}^{a_j}]$, where $n_t$ is the number of words. The objective is to derive a representation for each word.

Traditional word embedding models learn a single global vector for each word, which cannot capture context-dependent semantics when the same word appears in different descriptions. To address this limitation, we adopt BERT, a pre-trained language model built on multiple Transformer layers\cite{PapadopoulosPKN22}\cite{NIPS2017_3f5ee243}. Transformer models excel at capturing long-range dependencies in input sequences and directly relate words with arbitrary spacing \cite{zhang2023efficient}. This capability stems from the self-attention mechanism. For each token, the model computes attention weights over all other tokens by comparing learned query and key representations, then produces a
contextualized representation as a weighted combination of value vectors,
enabling it to aggregate relevant information regardless of positional
distance. BERT further incorporates positional embeddings to encode word order and capture spacing semantics.

Based on the above characteristics, WARBERT uses BERT to generate contextual embedding $\widehat{V}$ of dimension $F$. 
For the word sequence of the mashup $m_i$ description, we use a special format for input to the model:
\begin{align}
\widehat{V_r} = BERT([CLS], D^{m_i}, [SEP]),
\end{align}

where [SEP] is a special segmentation token and [CLS] is the classification token. 
Through BERT, the model learns to obtain the focus in the whole sequence.

\subsubsection{Classifier Utilizing Dual-Component Feature Fusion}
Standard BERT-based classifiers typically rely on either the pooler-output
or mean-pooling embeddings alone. However, these two representations capture
complementary information: the pooler-output encodes global sentence-level
signals via the [CLS] token, while mean-pooling preserves fine-grained
lexical details across all tokens. To leverage both and obtain richer semantic
representations, we propose a dual-component 
feature fusion strategy.
Once the contextual embedding $\widehat{V_r}$ is obtained, we derive the semantic feature embeddings for the mashup by leveraging the pooler-output of $\widehat{V_r}$ and applying mean-pooling to $\widehat{V_r}$. These two feature embeddings are then individually passed through two fully connected layers $FC_p$ and $FC_m$.
\begin{align}
    U_p = W_p \cdot Pooler(\widehat{V_r}) + b_p,
\end{align}
\begin{align}
    U_m = W_m\cdot MeanPooling(\widehat{V_r}) + b_m,
\end{align}
where $W \in \mathbb{R}^{L\times F}$ and $b \in \mathbb{R}^{L}$ are learnable parameters, with $L$ representing the number of APIs.

A fusion layer $FC_f$ is then employed to generate the API relevance score vector $v_r$ after concatenating $U_p$ with $U_m$. The vector $v_r$ is determined following the application of a sigmoid activation function.
\begin{align}
    v_r= sigmoid(W_f \cdot[U_p;U_m] + b_f),
\end{align}
where $W_f \in \mathbb{R}^{L\times 2L}$ and $b_f \in \mathbb{R}^{L}$ are learnable parameters.
Furthermore, by substituting the number of APIs with the number of API categories in the fully connected layers $FC_p$, $FC_m$,  and $FC_f$, WARBERT(R) can also judge the category of the API that the mashup may utilize.

\subsection{Implementation of WARBERT(M)}
\subsubsection{Attention Comparison}
A straightforward approach to measuring mashup-API similarity is to encode
their descriptions independently and compute similarity over the resulting
embeddings. However, it prevents the model from capturing token-level
interactions between the two texts during encoding. Instead, we concatenate
the mashup and API descriptions as a single input sequence, enabling the model to dynamically highlight functionally
relevant terms in the API description conditioned on the mashup
requirements. For example, when matching Yelp with a social API like
Twitter, the word ``reviews'' receives higher attention weight. 
For the input word sequence of the  API $a_j^{'}$ and mashup $m_i$, we use a special format:
\begin{align}
\widehat{V_m} = BERT([CLS], D^{m_i}, 
     [SEP],\allowbreak D^{a_j^{'}}, [SEP]).
\end{align}
Here, the self-attention mechanism is applied not only within each individual word sequence but also across the compared word sequences. In this cross-encoder setting, the [CLS] token already attends
to all tokens from both sequences and is thus well suited for capturing
their pairwise relationship, whereas mean-pooling would indiscriminately
mix tokens from both sequences.
As a result, in the API task layer $FC_t$, we solely utilize the pooler-output of $\widehat{V_m}$ as the feature embedding to get the similarity score between $m_i$ and $a_j^{'}$
:\begin{align}
    r^{m_i}_{j^{'}} = sigmoid(W_t \cdot  Pooler(\widehat{V_m}) + b_t).
\end{align}
By applying this methodology on each candidate API in $G^{m_i}=\{{a_1^{'}}, \ldots,{a_j^{'}},\ldots, {a_H^{'}}\}$, the mashup API similarity score vector $v_m$ is obtained:
\begin{align}
v_m = [r^{m_i}_{1^{'}},\ldots, r^{m_i}_{j^{'}}, \ldots r^{m_i}_{H^{'}}].
\end{align}
\subsection{Hierarchical Architecture with WARBERT(R) and WARBERT(M)}

The hierarchical architecture is designed to balance computational efficiency 
and semantic fidelity. Directly applying WARBERT(M) to the entire API repository 
would require $O(L)$ pairwise comparisons, which is computationally prohibitive. 
Moreover, without pre-filtering, WARBERT(M) is prone to false positives among 
semantically distant APIs. Conversely, WARBERT(R) alone lacks the fine-grained 
discriminative power to distinguish among semantically similar candidates. 
To address this, we employ WARBERT(R) as a filter to reduce the 
candidate space from $L$ to $H$ (where $H \ll L$), and WARBERT(M) as a 
 matcher to perform precise semantic comparison on this reduced set.
WARBERT(R) generate $H$ candidate APIs with high relevance scores  and WARBERT(M) to match these  candidates with the mashup to identify the final matched APIs. The final likelihood of whether an API matches a mashup is determined by the match score vector
$v$:\begin{align}
    v = \lambda*v_m+(1-\lambda)*v_r,
\end{align}
where $v_r$ is the API relevance score vector predicted by WARBERT(R), $v_m$ is the mashup API similarity score vector predicted by WARBERT(M) and $\lambda$ is a weighting factor ranging from 0 to 1. A higher
$\lambda$ places more emphasis on fine-grained semantic matching. To maintain dimensional consistency, only the relevance score vector of the $H$ candidate APIs is utilized.

\subsection{Training Process}
For the task of Web API recommendation, WARBERT is trained with a training set where APIs called by the mashup are labeled. For the auxiliary task of mashup category judgment, WARBERT(R) is trained separately using the categories of the labeled mashups. WARBERT uses the binary cross entropy function to compute the classification loss in both tasks. During training WARBERT converges to a state where it fits the dataset well via backpropagation.

\RestyleAlgo{ruled}
\LinesNumbered
\SetAlgoVlined
\IncMargin{0.5em}
\setlength{\textfloatsep}{0.2em}
\begin{algorithm} [t]
\caption{Computation of WARBERT}\label{alg:WARBERT}
  \SetKwInOut{Input}{Input} \SetKwInOut{Output}{Output}
  \Input{The description $D^{m_i}$ of the mashup $m_i$ , API repository  $A = [{a_1}, \ldots,{a_j},\ldots,{a_L}]$, the number of APIs $L$, API descriptions $[D^{a_1}, \ldots,D^{a_j},\ldots, D^{a_L}]$,  and the BERT model}
  \Output{a set of recommended APIs $S^{m_i}$}
  \For{mashup description $D^{m_i}$}{
    Apply WARBERT(R)\;
    \tcp{Compute Contextual Embedding}
    $\widehat{V_r} = BERT([CLS], D^{m_i}, [SEP])$\;
    \tcp{Extract Feature Vectors}
    $U_p = W_p \cdot Pooler(\widehat{V_r}) + b_p$\;
    $U_m = W_m \cdot MeanPooling(\widehat{V_r}) + b_m$\;
    
    \tcp{Get API Relevance Score Vector}
    $v_r = sigmoid(W_f \cdot [U_p; U_m] + b_f)$\;
  }
  \BlankLine
  rank and select the top-$H$ candidate APIs based on $v_r$\;
  fetch candidate APIs $G^{m_i}=\{{a_1^{'}}, \ldots,{a_j^{'}},\ldots, {a_H^{'}}\}$;
  \BlankLine
  
  \For{each candidate API ${a_j^{'}}$}{
    Apply WARBERT(M)\;
    \tcp{Compute Contextual Embedding}
  $\widehat{V_m} = BERT([CLS], D^{m_i}, 
     [SEP],\allowbreak D^{a_j^{'}}, [SEP])$\;
    \tcp{Calculate Similarity Score }
    $ r^{m_i}_{j^{'}}= sigmoid(W_t \cdot Pooler(\widehat{V_m}) + b_t)$\;
  }
  \BlankLine
  \tcp{Get Mashup API Similarity Score Vector}
  $v_m = [r^{m_i}_{1^{'}},\ldots, r^{m_i}_{j^{'}}, \ldots r^{m_i}_{H^{'}}]$;
  
  \tcp{Get the Final Match Score Vector}
  $v = \lambda \cdot v_m + (1 - \lambda) \cdot v_r$\;
  rank and select the top-$N$ APIs according to $v$\;
\Return $S^{m_i}$ = top-$N$ APIs\;
\end{algorithm}
In Algorithm~\ref{alg:WARBERT} we show the computation of the WARBERT model. WARBERT has both WARBERT(R) and WARBERT(M) implementations and the training procedures are similar:

1) Computing contextual embeddings using pre-trained BERT models.

2) Extracting feature vectors from the contextual embeddings.

3) The final matching results are obtained using the classifier that takes feature vectors as input and is trained on the training set.

 The parameters of WARBERT(R) and WARBERT(M) are initialized with the pre-trained BERT. They are fine-tuned by the training process described above.

\section{Experiments And Analysis}
\label{sec:expanalysis}
To evaluate the effectiveness of WARBERT, we need to answer the following research questions:

RQ1: How effective is WARBERT in recommending accurate Web APIs when compared with the baseline models?

RQ2: How effective is WARBERT(R) in the auxiliary task mashup category judgment compared to the model MTFM and MTFM++?

RQ3: What is the computational cost of WARBERT and the baseline models?

RQ4: How effective is the hierarchical architecture of WARBERT?

RQ5: How effective is the dual-component feature fusion strategy employed by WARBERT?

RQ6: How effective is the attention comparison strategy employed by WARBERT?

RQ7: What influence do the weighting factor $\lambda$ and the candidate API number $H$ have on WARBERT?

\subsection{Benchmark Dataset}

Following MTFM, we use the Mashup and API data crawled in ProgrammableWeb. Since the MTFM model requires data with three attributes: name, description and category. MTFM removes crawled data with missing attribute values. Finally, the dataset contains 8217 mashups and 1647 APIs. Each mashup has at least one API call, and on average 2 APIs are called.  53.82\% of all mashups call one API, 23.83\% call two APIs. The remaining 22.36\% require three or more APIs, of which 7.39\% involve five or more APIs. The ratio of positive pair (mashup called API) is 0.125\%. Detailed statistics of the dataset are presented in Table~\ref{tab:dataset}.

\begin{table}[t]
\caption{Statistics of the dataset}\label{tab:dataset}
\begin{center}
\begin{tabular}{l|c}
\toprule
\textbf{Statistics}& \textbf{Value}\\
\midrule
Number of APIs& 1647 \\
Number of Mashups& 8217\\
Number of Categories& 499\\
APIs per Mashup& 2.091\\
Categories per Mashup& 2.447\\
Categories per API& 2.271\\
Words in description per Mashup& 17.019\\
Words in description per API& 44.296\\
\bottomrule
\end{tabular}
\end{center}
\end{table}

\begin{table*}[t]
\caption{Performance comparison of baseline models on the task of Web API recommendation}\label{tab:result}
\centering
\begin{tabular}{l|cccc|cccc|cccc}
\toprule
\multirow{2}*{\textbf{Models}}& \multicolumn{4}{c|}{N = 1}& \multicolumn{4}{c|}{N = 5}& \multicolumn{4}{c}{N = 10}\\
~ & \textbf{Prec}& \textbf{Rec}& \textbf{NDCG}& \textbf{MAP}& \textbf{Prec}& \textbf{Rec}& \textbf{NDCG}& \textbf{MAP}& \textbf{Prec}& \textbf{Rec}& \textbf{NDCG}& \textbf{MAP}\\
\midrule
FC-LSTM& 0.3087& 0.2153& 0.3087& 0.3087& 0.1328& 0.3784& 0.4272& 0.3864& 0.0842& 0.4614& 0.4524& 0.3911\\
SPR& 0.4187& 0.2868& 0.4187& 0.4187& 0.1744& 0.4874& 0.5568& 0.5151& 0.1114& 0.5970& 0.5784& 0.5089\\
RWR& 0.4291& 0.3097& 0.4291& 0.4291& 0.1876& 0.5449& 0.5997& 0.5510& 0.1153& 0.6431& 0.6200& 0.5471\\
MTFM& 0.5898& 0.3939& 0.5898& 0.5898& 0.1920& 0.5535& 0.6613& 0.6324& 0.1111& 0.6066& 0.6644& 0.6174\\
MTFM++& 0.5861& 0.3957& 0.5861& 0.5861& 0.1940& 0.5518& 0.6595& 0.6306& 0.1136& 0.6127& 0.6637& 0.6164\\
ServiceBERT & 0.5806 & 0.4019 & 0.5806 & 0.5806 & 0.1680 & 0.5329 & 0.5275 & 0.4779 & 0.0925 & 0.5731 & 0.5396 & 0.4818 \\ 
SEHGN & 0.6173 & 0.4298 & 0.6173 & 0.6173 & 0.2060 & 0.6114 & 0.6381 & 0.6187 & 0.1042 & 0.6498 & 0.6407 & 0.6201 \\
\textbf{WARBERT}&\textbf {0.6616}& \textbf{0.4510}& \textbf{0.6616}& \textbf{0.6616}&\textbf{0.2167}& \textbf{0.6172}& \textbf{0.7173}&\textbf{0.6945}&\textbf{0.1215}& \textbf{0.6615}&\textbf{0.7189}& \textbf{0.6825}\\
\bottomrule
\end{tabular}
\end{table*}

\begin{table*}[t]
\caption{Performance comparison of baseline models on the task of mashup category judgment}\label{tab:result_c}
\centering
\begin{tabular*}{0.6\textwidth}{@{\extracolsep{\fill}}c|cc|cc|cc}
\toprule
\multirow{2}*{\textbf{Models}}& \multicolumn{2}{c|}{N = 1}& \multicolumn{2}{c|}{N = 5}& \multicolumn{2}{c}{N = 10}\\
~ & \textbf{Rec}& \textbf{NDCG}& \textbf{Rec}& \textbf{NDCG}& \textbf{Rec}& \textbf{NDCG}\\
\midrule
MTFM& 0.3276& 0.6439& 0.6000& 0.7446& 0.6934& 0.7484\\
MTFM++& 0.3384& 0.6598& 0.6017& 0.7553& 0.6904& 0.7571\\
\textbf{WARBERT(R)}& \textbf{0.3621}& \textbf{0.6945}& \textbf{0.6335}& \textbf{0.7862}& \textbf{0.7219}&\textbf{0.7850}\\
\bottomrule
\end{tabular*}
\end{table*}

\subsection{Baseline Models}

There is no universal benchmark for the API recommendation problem. Therefore, we use some representative methods as benchmarks. These include the previous solution FC-LSTM, SPR, RWR, MTFM and its variant MTFM++, ServiceBERT and SEHGN. We summarize these methods in the following.

\begin{itemize}
    \item \textbf{FC-LSTM}: FC-LSTM~\cite{shi2019functional} uses an additional attention mechanism based on the LSTM model. The attention mechanism is used to mine the function features in mashups and APIs. In addition, FC-LSTM also uses LDA models for text enhancement. To ensure fairness, we disable this strategy.  
    \item \textbf{SPR}: The Service Profile Reconstruction Model (SPR)~\cite{ZhongF0Z18} uses topic modeling to analyze the connection between mashup descriptions and APIs. It offers API recommendations based on the topic structure of the descriptions.
    \item \textbf{RWR}: The Random Walk with Restart (RWR) model~\cite{wang2018mashup} encodes mashup-specific contexts in a knowledge graph and ranks Web APIs by relevance using a schema that models mashup requirements with category entities. The RWR algorithm prioritizes these entities to estimate API relevance.
    \item \textbf{MTFM}: MTFM\footnote{The source code is at \url{https://github.com/whale-ynu/MTFM}}~\cite{wu2021mashup} is the model based on CNN. It uses a multi-layer CNN network to extract the features of the mashup and the API. It concatenates the mashup feature with all the API features to get the advantage of collaborative filtering. Furthermore, it combines multi-task learning, and uses mashup category judgment as an auxiliary task.
    \item \textbf{MTFM++}: MTFM++ is an extension of MTFM. It additionally uses the category attributes of the API and constructs quality features using Quality of Services (QoS) related attributes. Due to the additional use of attribute data, MTFM++ requires more demanding data formats.
   \item \textbf{ServiceBERT}: ServiceBERT~\cite{wang2021servicebert} levarages masked language modeling, replaced token detection objectives, and contrastive learning to pretrain BERT, enabling noise invariant sentence-level semantic representations of Mashups and APIs.   
    \item \textbf{SEHGN}: SEHGN~\cite{WangXLPWDY24} integrates a semantic embedding component to capture text semantics and a multi-semantic aggregator to encode multi-type structural relationships in the Mashup-API Heterogeneous Graph, addressing API compatibility dependencies and sparse invocation records.
    \item \textbf{WARBERT}: The WARBERT framework encompasses three implementations, namely WARBERT itself, along with its component WARBERT(R) and WARBERT(M). All implementations are evaluated on their performance in the Web API recommendation task. Furthermore WARBERT(R) shows the model's multi-task learning ability in the task of mashup category judgment.
    
\end{itemize}

\subsection{Experimental Settings}

\textbf{Dataset configuration.}
We directly adopt the preprocessed dataset released by Wu \etal~\cite{wu2021mashup},
which already applies text filtering, abbreviation replacement, and lemmatization to
mashup and API descriptions.
The dataset is split into training, validation, and test sets with a ratio of
3:1:1 (random seed = 42), and all methods are evaluated on identical splits
for fair comparison.

\textbf{Training configuration.}
All experiments are conducted on a single NVIDIA L20 (48\,GB) GPU. WARBERT adopts BERT-Tiny~\cite{bhargava2021generalization,DBLP:journals/corr/abs-1908-08962} as the backbone, consisting of 2
Transformer layers, 2 attention heads, and a hidden size of 128,
totaling approximately 4.4M parameters.
WARBERT(R) is trained on the full training dataset for approximately 15 epochs, with the learning rate set to
$1\times10^{-3}$ for the first 6 epochs and $1\times10^{-5}$ for the
remaining epochs. For WARBERT(M), training pairs are constructed by pairing
each mashup in the training set with its ground-truth APIs as positives and 50 negative APIs per
positive instance, of which 60\% are hard negatives (APIs ranked within the
top-$H$ by WARBERT(R) but absent from the ground truth) to help the
model learn to distinguish borderline cases and the remaining
40\% are randomly sampled to maintain diversity. WARBERT(M) is trained for approximately 20 epochs
with the learning rate starting at $1\times10^{-3}$ for the first 16 epochs
and $1\times10^{-5}$ for the subsequent epochs. Both components use the Adam
optimizer with a batch size of 64 and a maximum input sequence length of 256.
Each epoch is validated on the validation set to monitor overfitting, and an
early stopping mechanism with patience 7 is employed. The model achieving the
highest validation score is selected for testing. The candidate number $H$ is
set to 45 and the weighting factor $\lambda$ is set to 0.6, which yielded
the best results.

\textbf{Baseline configuration.}
For baseline methods, we adopt the hyperparameter settings reported in their
original publications. For parameters not explicitly specified, we use commonly
recommended configurations from related literature. Specifically, for SEHGN we
set the patience to 7 and the maximum epochs to 100 following MTFM. For
ServiceBERT, we set the fine-tuning learning rate to $5\times10^{-5}$ and the
batch size to 32.

\subsection{Evaluation Metrics}

As our experiment is to judge whether the mashup and API match, we pay more attention to the quality of the correct recommended APIs. We use Precision and Recall as the common evaluation metric. Usually, users pay more attention to APIs with high match scores in the recommendation list. To follow other methods, we also use Top-$N$ results to evaluate the model. We use $N$ as 1, 5 and 10. 

Precision@N refers to the ratio of the number of real hits in the
recommended APIs to the number of recommended APIs.
\begin{align}
    Precision@N = \frac{|\text{{APIs}}_\text{{real}} \cap \text{APIs}_\text{top-N }|}{| \text{APIs}_\text{top-N}|}.
\end{align}
Recall@N refers to the ratio of the number of real hits in the
recommended APIs to the number of all real APIs.
\begin{align}
    Recall@N = \frac{|\text{APIs}_\text{real} \cap \text{APIs}_\text{top-N}|}{|\text{APIs}_\text{real}|}.
\end{align}

Higher values for these metrics indicate greater accuracy and enhanced performance of the model.
Also, since Precision@N and Recall@N do not take into account the order in the results. We used additional evaluation metrics: MAP@N and NDCG@N. The Mean Average Precision (MAP) and Normalized Discounted Cumulative Gain (NDCG) are commonly used evaluation metrics in information retrieval tasks, and they consider the order in the results. NDCG is calculated from DCG by normalizing IDCG. IDCG is the ideal DCG. DCG takes into account the ranking order, so that the top ranked items gain more and the bottom ranked items are discounted.
\begin{gather}
    DCG@N = \sum_{i=1}^N\frac{rel_i}{log_2(i + 1)}, \\
    IDCG@N = \sum_{i=1}^{|real APIs|}\frac{1}{log_2(i + 1)}, \\
    NDCG@N = \frac{DCG@N}{IDCG@N},
\end{gather}
where $rel_i = 1 / 0$ represents whether the $i$-th API is truly relevant to the current mashup or not. AP calculates Precision at each position, and averages over all matching positions.
\begin{align}
    AP@N = \frac{\sum_{i=1}^N rel_i * Precision@i}{\sum_{i=1}^N rel_i}.
\end{align}
 MAP is obtained by averaging the Average Precision (AP) across all mashups $M$:
 \begin{align}
    MAP@N = \frac{1}{|M|} \sum_{m \in M} AP_m@N.
\end{align}

\subsection{Main Results}
\noindent\textit{1) RQ1: Performance of WARBERT on Web API recommendation.}

WARBERT outperforms the other models across all metrics. Table~\ref{tab:result} shows the results of the performance on the task of Web API recommendation. WARBERT outperforms the best-performing baseline SEHGN in all metrics. For Precision and Recall, WARBERT improves by up to 16.60\% and 4.93\%. For NDCG and MAP, WARBERT averages 10.60\% and 9.83\% improvements over SEHGN.

FC-LSTM gives higher weight to the function information in the description to improve the effect. However, FC-LSTM treats the API function information as composed of words semantically similar to the tag. So FC-LSTM needs additional tag information. Due to data heterogeneity, the tag information in the dataset may not be accurate. In WARBERT, the descriptions of the mashup and the API are compared directly. During the comparison process, WARBERT learns keywords, leading to improved performance.

SPR outperforms FC-LSTM. Its hidden topic layer addresses the limitations of the bag-of-words model. It enables semantic connections between requirements and APIs even when documents are sparse. RWR  surpasses SPR probably due to its more descriptive tag composition, which better captures developer requirements. However, it also raises the operational complexity for users who must accurately and comprehensively articulate their needs.

For MTFM, it combines the advantages of collaborative filtering and multi-task learning. This makes its results better than RWR. However, also due to the requirement of multi-task learning, MTFM needs to contain labeled data in the dataset. This makes it more difficult for developers to operate. MTFM++ further imposes additional requirements on the data. The API is required with the corresponding QoS information to obtain the quality vector. Finally, in MTFM FIC (Feature Interaction Component), the mashup embedding and all API embeddings are concatenated together into a single embedding. 

ServiceBERT uses BERT to gain enriched semantics. However, using cosine similarity to match API and mashup descriptions neglects the differences in natural language usage between service users and developers, thus leading to poorer performance than WARBERT.

SEHGN captures text semantics via a semantic embedding component and encodes structural relationships in a heterogeneous graph. However, it relies on graph propagation, which may struggle with semantic ambiguities due to less precise embeddings. Moreover, it exhibits reduced performance on small datasets, as it requires sufficient data to capture long-range dependencies effectively.

Compared to other baseline models,  WARBERT has significant improvements in all evaluation metrics. WARBERT employs a lightweight variant of BERT for fine-tuning, implying that scaling the language model's capacity could improve performance. Moreover, WARBERT does not require additional information, making it more general and suitable for actual scenarios. This also demonstrates the potential of WARBERT to improve further by enriching its information sources. 

\noindent\textit{2) RQ2: Performance of WARBERT on mashup category judgement.} 

WARBERT(R) performs better than MTFM and MTFM++.
Table~\ref{tab:result_c} shows the results of the performance on the task of mashup category judgement. On average, WARBERT(R) demonstrates a performance improvement of approximately 5.6\% in Recall and 4.4\% in NDCG compared to MTFM++.

\begin{table*}[t]
\caption{Performance comparison of ablation models on the task of Web API recommendation}\label{tab:a_result}
\centering
\begin{tabular}{l|cc|cccc|cccc}
\toprule
\multirow{2}*{\textbf{Models}}& \multicolumn{2}{c|}{N = 1}& \multicolumn{4}{c|}{N = 5}& \multicolumn{4}{c}{N = 10}\\
~ & \textbf{Prec}& \textbf{Rec}& \textbf{Prec}& \textbf{Rec}& \textbf{NDCG}& \textbf{MAP}& \textbf{Prec}& \textbf{Rec}& \textbf{NDCG}& \textbf{MAP}\\
\midrule

WARBERT(w/o meanpooling)& 0.5928& 0.3965& 0.1850& 0.5116& 0.6381& 0.6198& 0.1037& 0.5496& 0.6432& 0.6133\\

WARBERT(w/o pooler-output)& 0.6281& 0.4235& 0.2050& 0.5754& 0.6829& 0.6597& 0.1155& 0.6186& 0.6856& 0.6486\\

WARBERT(w/o attention)& 0.6360& 0.4316& 0.2133& 0.6094& 0.7043& 0.6786& 0.1202& 0.6541& 0.7080& 0.6683\\
  WARBERT(M)& 0.6147& 0.4170& 0.1939& 0.5575& 0.6657& 0.6460& 0.1096& 0.6000& 0.6707& 0.6372\\
  WARBERT(R)& 0.6379& 0.4295& 0.2100& 0.6016& 0.6995& 0.6749& 0.1180& 0.6463& 0.7021& 0.6636\\
\textbf{WARBERT}& \textbf{0.6616}& \textbf{0.4510} &\textbf{0.2167}& \textbf{0.6172}& \textbf{0.7173}& \textbf{0.6945}& \textbf{0.1215}& \textbf{0.6615}& \textbf{0.7189}& \textbf{0.6825}\\

\bottomrule
\end{tabular}
\end{table*}

\begin{table*}[t]
\caption{Performance comparison of ablation models on the task of mashup category judgment}\label{tab:a_result_c}
\centering
\begin{tabular*}{0.7\textwidth}{@{\extracolsep{\fill}}l|cc|cc|cc}
\toprule
\multirow{2}*{\textbf{Models}}& \multicolumn{2}{c|}{N = 1}& \multicolumn{2}{c|}{N = 5}& \multicolumn{2}{c}{N = 10}\\
~ & \textbf{Rec}& \textbf{NDCG}& \textbf{Rec}& \textbf{NDCG}& \textbf{Rec}& \textbf{NDCG}\\
\midrule
WARBERT(R w/o meanpooling)& 0.1819& 0.3682& 0.3850& 0.5176& 0.5154& 0.5455\\
WARBERT(R w/o pooler-output)& 0.3474& 0.6701& 0.6120& 0.7638& 0.6927& 0.7644\\

\textbf{WARBERT(R)}& \textbf{0.3621}& \textbf{0.6945}& \textbf{0.6335}& \textbf{0.7862}& \textbf{0.7219}& \textbf{0.7850}\\

\bottomrule
\end{tabular*}
\end{table*}

\noindent \textit{3) RQ3: Computational cost of WARBERT.}
\newline
Table~\ref{tab:cost} reports the training time of each model to attain its best performance, the inference time for each mashup, and the peak GPU memory during training.
Compared to match-type baselines such as FC-LSTM and ServiceBERT, WARBERT has shorter inference time due to efficient pair-to-pair matching. Compared to recommendation-type solutions, WARBERT takes longer time for inference but achieves superior performance. Its training time (1.1 minutes for WARBERT(R) and 3.5 hours for WARBERT(M)) is longer than lightweight methods such as MTFM but comparable to ServiceBERT. Using a lightweight BERT-Tiny backbone, WARBERT requires substantially less GPU memory, making it viable for real-world deployment on resource-constrained environments.

\begin{table}[t]
\caption{Computational cost of WARBERT and the baselines}\label{tab:cost} 
\centering

\begin{tabular}{l|c|c|c}
\toprule
\textbf{Methods} & \textbf{Training Time} & \textbf{Inference Time} & \textbf{GPU Memory} \\
\midrule
FC-LSTM & 15.1 m & 2.5 s & 0.9 GB \\
SPR & 2.0 m & 0.4 ms & 0.7 GB \\
RWR & -- & 26.7 ms & -- \\
MTFM & 3.3 m & 0.6 ms & 0.5 GB \\
ServiceBERT & 4.5 h & 7.8 s & 42.7 GB \\
SEHGN & 1.4 h & 3.7 ms & 1.5 GB \\
WARBERT & 1.1 m+ 3.5 h & 0.5 s & 0.6 GB \\
\bottomrule
\end{tabular}
\end{table}

\subsection{Ablation Study}

To answer RQ4-RQ7, we design the ablation models and conduct the ablation study.

\noindent\textit{4) RQ4: Effectiveness of the Hierarchical Architecture of WARBERT.}

WARBERT demonstrates significant performance improvements over its individual components. Specifically, for Precision and Recall, WARBERT shows enhancements of up to 10.1\% and 9.7\% respectively compared to WARBERT(M). Additionally, for NDCG and MAP, WARBERT achieves average improvements of 7.5\% and 7.4\% over WARBERT(M). When compared to WARBERT(R), WARBERT's Precision and Recall see improvements of up to 3.3\% and 3.3\%, with average enhancements for NDCG and MAP at 2.9\% and 3.1\%. These results empirically validate the complementary design of the hierarchical 
architecture. When used alone, WARBERT(R) achieves broad candidate coverage 
but struggles to distinguish among semantically similar APIs, leading to 
suboptimal ranking quality. Conversely, WARBERT(M) alone exhibits lower 
recall, as it lacks an effective filtering mechanism and may miss relevant 
APIs that do not rank highly in initial retrieval. By combining the two 
stages, WARBERT leverages the efficiency of WARBERT(R) for candidate 
generation and the discriminative power of WARBERT(M) for fine-grained 
semantic matching, resulting in consistently superior performance across 
all evaluation metrics.

\noindent\textit{5) RQ5: Effectiveness of Dual-component Feature Fusion.}

 In WARBERT, the performance is significantly reduced when the dual-component feature fusion is replaced with utilizing a single feature. 
 Specifically, by removing the feature vector from the pooler-output of the contextual embedding, we obtain WARBERT(R w/o pooler-output), and by removing the feature vector from the Mean-Pooling of the contextual embedding, we obtain WARBERT(R w/o meanpooling). Moreover, by substituting WARBERT(R) in the original WARBERT with WARBERT(R w/o pooler-output) or WARBERT(R w/o meanpooling), we obtain WARBERT(w/o pooler-output) and WARBERT(w/o meanpooling) respectively.
In the task of API recommendation, a comparison between WARBERT(R) and WARBERT(R w/o pooler-output) reveals a performance decrease ranging from 7.2\% to 9.0\% across all evaluation metrics. In contrast, when comparing WARBERT(R) with WARBERT(R w/o meanpooling), a more substantial performance decline is observed, ranging from 38.0\% to 44.5\%. Additionally, the comparison between WARBERT(w/o pooler-output) and WARBERT shows a reduction in performance of 4.6\% to 6.5\%, while comparing WARBERT(w/o meanpooling) with WARBERT results in a more significant decrease of 10.1\% to 17.1\%.
Similarly, in the task of mashup category judgment, WARBERT(R w/o pooler-output) experiences a reduction of 2.6\% to 4.1\%, while WARBERT(R w/o meanpooling) sees a more significant drop of 28.6\% to 49.8\%. These results indicate that the dual-component feature fusion enhances model performance by providing more comprehensive semantics.

\noindent\textit{6) RQ6: Effectiveness of Attention Comparison.}

In WARBERT the performance is greatly reduced when the attention comparison is replaced with a simplified encoding concatenation approach.
We use BERT-Tiny to encode mashup descriptions and API descriptions separately. We then concatenate their pooler-output and feed them into the API task layer to get the similarity score vector. The variant is denoted as WARBERT(M w/o attention), and by substituting WARBERT(M) in the original WARBERT, we derive WARBERT(w/o attention). Relative to WARBERT(M), WARBERT(M w/o attention) exhibits a performance reduction ranging from 20.7\% to 48.5\%, whereas WARBERT(w/o attention) shows a decrease of between 1.1\% and 4.3\% compared to WARBERT. These results demonstrate that jointly encoding mashup and API
descriptions enables the attention mechanism to prioritize
functionally relevant terms, reducing semantic discrepancies
between user-oriented mashup descriptions and developer-provided
API documentation.

\begin{figure}[!t]
\centering
\includegraphics[width=\linewidth]{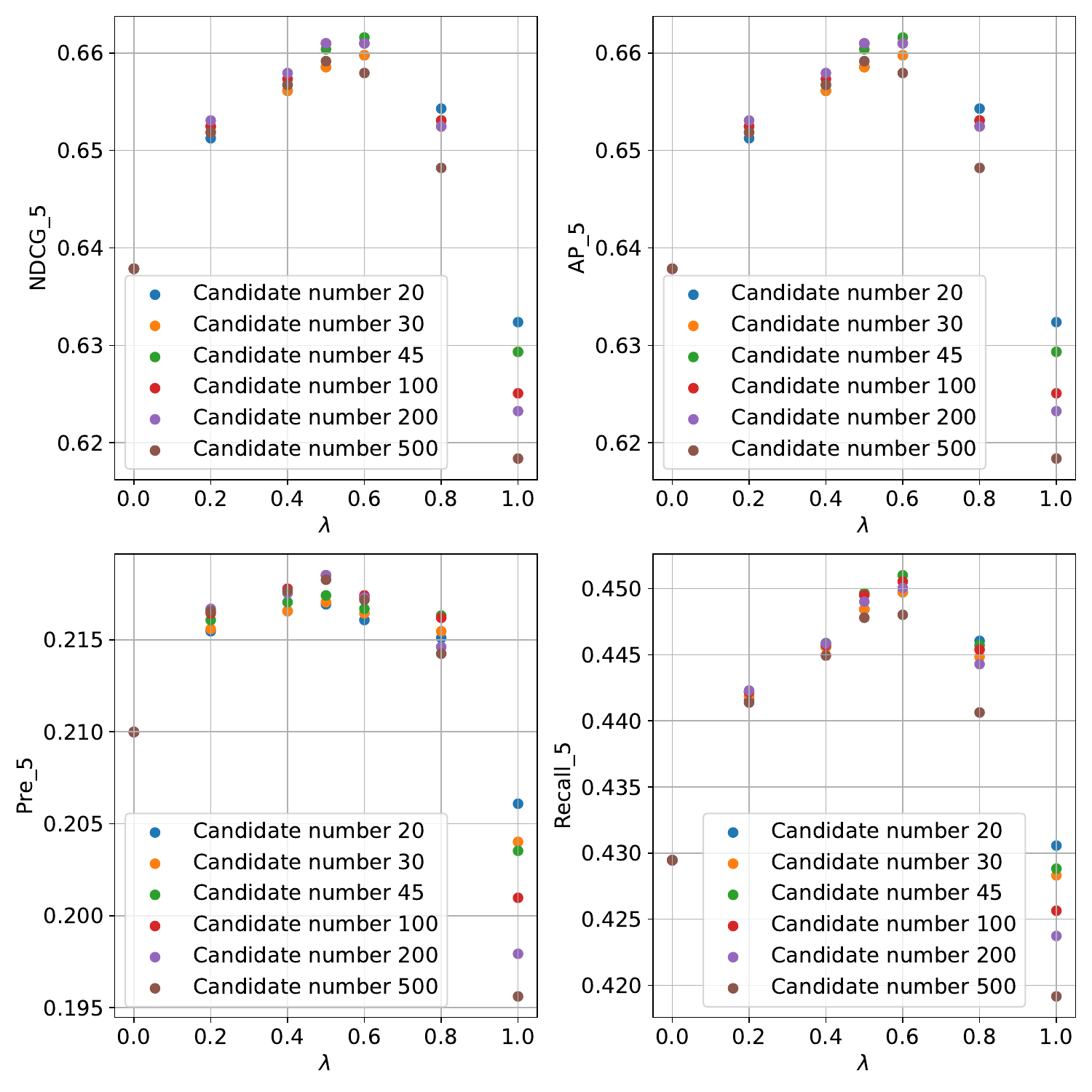}
\caption{Performance of WARBERT with different candidate number $H$ and weighting factor $\lambda$}
\label{fig:metric}
\end{figure}

\begin{figure}[!t]
\centering
\includegraphics[width=\linewidth]{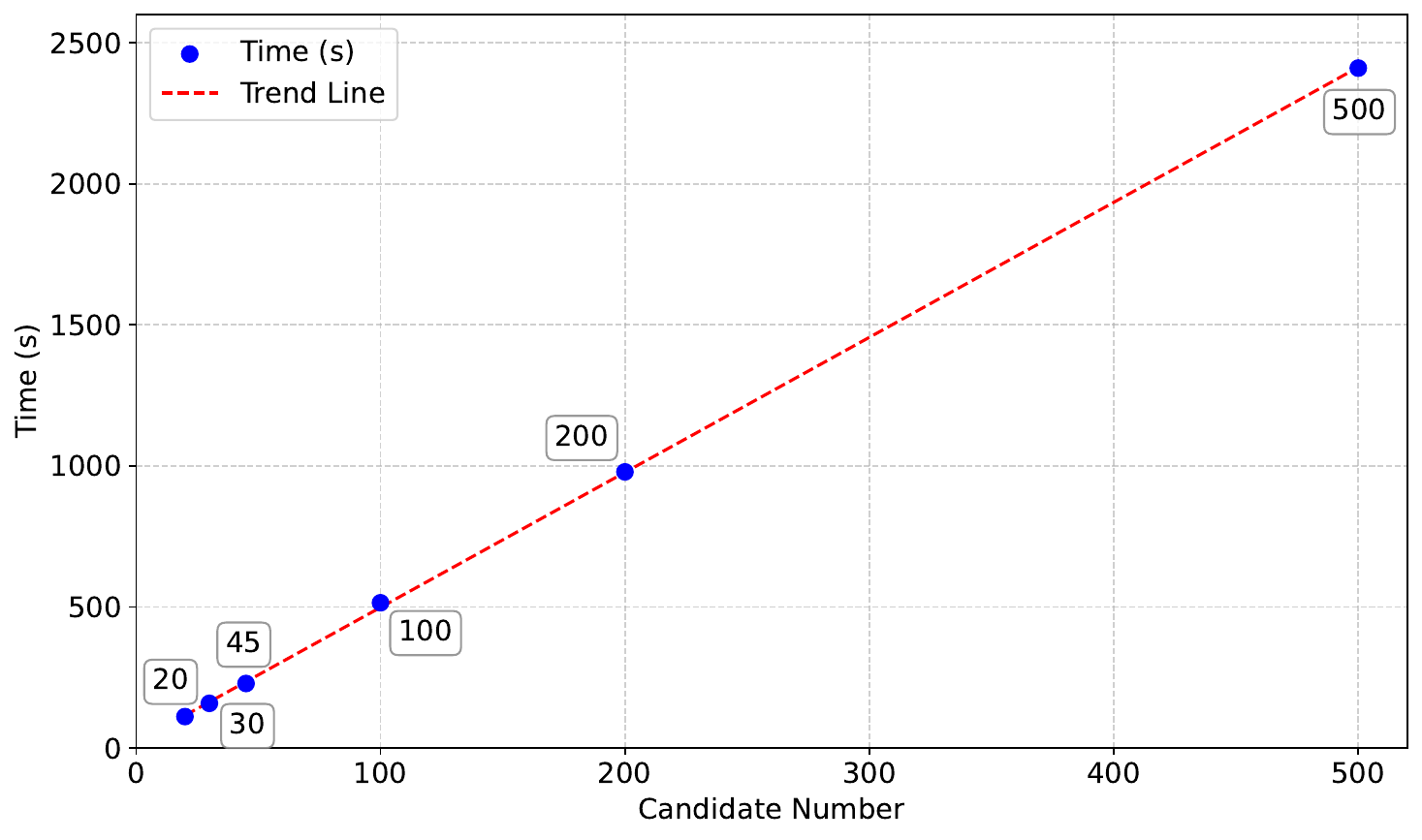}
\caption{Time consumption for WARBERT with different candidate number $H$}
\label{fig:time}
\end{figure}

\begin{table*}[!t]
\caption{Case-Study Analyses}
\label{tab:case_study}
\small
\begin{tabular}{p{2.5cm} l}
\toprule 
\multicolumn{2}{l}{\textbf{Mashup:}  egypt-forecast \textbf{Used APIs: \textbf{world-weather-online}, \textbf{openweathermap}, \textbf{google-maps}, \textbf{google-chart}, \textbf{panoramio}} } \\
\multicolumn{2}{p{18cm}}{\textbf{Description}: \textit{Egypt Forecast provides weather conditions for 70 Egyptian cities using Google Maps and different  
  weather APIs.  }} \\ \hline
Method & Generated APIs \\ \hline
FC-LSTM & \textbf{google-maps}, facebook, twitter, youtube, weather-channel\\
SPR &  \textbf{google-maps}, twitter, youtube, flickr, facebook \\
RWR & \textbf{google-maps}, mapbox, cicero, geograph, uber\\
MTFM & google-geocoding, \textbf{google-maps}, forecast, \textbf{panoramio}, \textbf{world-weather-online} \\
MTFM++ & \textbf{google-maps}, weatherbug, \textbf{google-chart}, yahoo-weather, geonames \\
ServiceBERT & \textbf{google-maps}, geonames, \textbf{google-chart} , yahoo-maps, \textbf{world-weather-online} \\
SEHGN & \textbf{google-maps}, \textbf{panoramio}, google-geocoding, google-maps-places, \textbf{google-chart} \\
\textbf{WARBERT} & \textbf{panoramio}, \textbf{google-maps}, \textbf{google-chart}, \textbf{world-weather-online}, weatherbug \\ 
\hline\hline

\multicolumn{2}{l}{\textbf{Mashup:} recordlective \textbf{Used APIs: \textbf{youtube}, \textbf{last.fm}, \textbf{musicbrainz}}} \\
\multicolumn{2}{p{18cm}}{\textbf{Description}:  \textit{Recordlective combines various sources to provide music fans with the ability to stream music albums from start to finish.}} \\ \hline
Method & Generated APIs \\ \hline
FC-LSTM & \textbf{youtube}, google-maps, facebook, twitter, flickr \\
SPR & \textbf{youtube}, google-maps, \textbf{last.fm}, twitter, flickr\\
RWR & \textbf{last.fm}, \textbf{musicbrainz}, google-maps, twitter, flickr \\
MTFM & \textbf{youtube}, \textbf{musicbrainz}, lyricsfly, twitter, bandsintown \\
MTFM++ &  \textbf{youtube},  \textbf{last.fm}, google-maps, soundcloud, twitter \\
ServiceBERT &  \textbf{youtube}, \textbf{last.fm}, exfm, google-talk, myvox\\
SEHGN & \textbf{youtube}, \textbf{last.fm}, spotify-metadata, \textbf{musicbrainz}, lyricwiki \\
\textbf{WARBERT} & \textbf{last.fm}, \textbf{youtube}, \textbf{musicbrainz}, twitter, amazon-product-advertising \\ 
\hline\hline

\multicolumn{2}{l}{\textbf{Mashup:} Ohsopopular \textbf{Used APIs: del.icio.us, google-search, alexa-web-information-service, compete, backtype, bing}} \\
\multicolumn{2}{p{18cm}}{\textbf{Description}: \textit{Ohsopopular is a service to see how popular a URL is on the web. Checks several services like Google PageRank, Alexa Rank, Compete, Delicious, etc., and assigns a number rank.}}\\ \hline
Method & Generated APIs \\ \hline
FC-LSTM & twitter, youtube, flickr, \textbf{bing}, \textbf{backtype} \\
SPR & google-maps, twitter, youtube, flickr, facebook \\
RWR &  \textbf{backtype}, google-maps, twitter, youtube, flickr \\
MTFM & \textbf{del.icio.us}, \textbf{google-search}, cnet, google-app-engine, box \\
MTFM++ & \textbf{del.icio.us}, google-adsense, \textbf{backtype}, disqus, \textbf{google-search}\\
ServiceBERT & \textbf{bing}, microsoft-bing, yahoo-search, pinboard, shadows \\
SEHGN &\textbf{google-search}, \textbf{alexa-web-information-service}, \textbf{del.icio.us}, yahoo-search, crunchbase \\
\textbf{WARBERT} & \textbf{bing}, \textbf{alexa-web-information-service}, \textbf{del.icio.us}, \textbf{google-search}, twitter \\ 
\hline\hline

\multicolumn{2}{l}{\textbf{Mashup:} ebay-nintendo-wii-ticker \textbf{Used APIs: google-desktop, ebay}}\\
\multicolumn{2}{p{18cm}}{\textbf{Description}: \textit{Tired of hunting elusive Nintendo Wiis. Use this Google Gadget as a live ticker displaying Nintendo
   Wiis for sale on your chosen eBay site. Supports US, UK, and Australian eBay sites. Use it to grab
   a bargain.}} \\ \hline
Method & Generated APIs \\ \hline
FC-LSTM & twitter, youtube, flickr, \textbf{ebay}, google-search \\
SPR & google-maps, \textbf{ebay}, twitter, youtube, flickr \\
RWR & google-maps, twitter, youtube, flickr, \textbf{ebay} \\
MTFM & \textbf{ebay}, google-homepage, google-search, google-calendar, microsoft-bing-maps \\
MTFM++ & \textbf{ebay}, google-search, paypal, channeladvisor, dataunison-ebay-research \\
ServiceBERT & google-homepage, google-base, \textbf{ebay}, google-search, ebay-finding \\
SEHGN &  google-calendar, facebook, google-cloud-translation, paypal, \textbf{ebay} \\
\textbf{WARBERT} & \textbf{ebay}, google-maps, amazon-product-advertising, google-homepage, \textbf{google-desktop} \\
\bottomrule
\multicolumn{2}{l}{The right APIs are marked in bold.}
\end{tabular}

\end{table*}

\noindent\textit{7) RQ7: Influence of $\lambda$ and the candidate number $H$.} 

We analyze the sensitivity of WARBERT to two key hyperparameters: the candidate number $H$ and the weighting factor $\lambda$. Figure~\ref{fig:metric} shows the results with $H\in\{20, 30, 45, 100, 200, 500\}$ and $\lambda \in \{0, 0.2, 0.4, 0.5, 0.6, 0.8, 1.0\}$ using Top-5 results. The weighting factor $\lambda$ balances the contributions of WARBERT(R) and WARBERT(M). The results show that all metrics except Precision peak at $\lambda$=0.6, indicating that integrating both components yields better performance than using either alone. Notably, even when $\lambda$=1, the hierarchical framework still outperforms standalone WARBERT(M) without filtering, demonstrating the effectiveness of WARBERT(R) as a candidate filter. The candidate number $H$ controls the size of the candidate pool passed from WARBERT(R) to WARBERT(M). At $\lambda$=0.6, all metrics except Precision peak at $H$=45, which accounts for approximately 2.7\% of the entire API repository. When $H$ is too small, relevant APIs may be incorrectly filtered out; when $H$ is too large, increased noise degrades matching accuracy. Furthermore, as shown in Figure~\ref{fig:time}, $H$ exhibits an approximately linear positive correlation with inference time, confirming the efficiency benefits of the hierarchical filtering strategy. Overall, WARBERT demonstrates robust performance across a reasonable range of parameter values, with variation across all metrics remaining within 2.0\% for $\lambda \in [0.2, 0.8]$ and $H \in [45, 200]$.

\subsection{Case Study}
To intuitively demonstrate the recommendation quality of WARBERT, we select four representative mashups from the test set covering diverse application domains and present the top-5 recommended APIs of each method in Table~\ref{tab:case_study}. In the case of \textbf{egypt-forecast}, WARBERT successfully retrieves four of five ground-truth APIs, outperforming SEHGN (three hits) and ServiceBERT (three hits). In contrast, earlier methods such as FC-LSTM and SPR tend to recommend generic social APIs that are irrelevant to weather or mapping functionalities. Notably, WARBERT(R) alone retrieves google-maps, panoramio, and google-chart, while WARBERT(M) captures panoramio and google-maps but introduces semantically distant APIs such as flickr and wikipedia. 
For \textbf{recordlective}, WARBERT recovers all three ground-truth APIs within the top-3 positions. SEHGN also achieves full recall but includes extraneous candidates such as spotify-metadata. Baselines like FC-LSTM and RWR suggest unrelated APIs such as google-maps. An examination of the individual components reveals that WARBERT(R) retrieves all three APIs but ranks musicbrainz below lyricwiki, whereas WARBERT(M) places amazon-product-advertising and twitter ahead of musicbrainz. Regarding \textbf{Ohsopopular},  WARBERT correctly identifies four of six ground-truth APIs, outperforming SEHGN (three hits) and ServiceBERT (one hit). Other baselines return largely irrelevant results. However, WARBERT misses the less common APIs compete and backtype, likely because their short and generic descriptions provide insufficient semantic signals for accurate matching. For \textbf{ebay-nintendo-wii-ticker}, WARBERT is the only method that retrieves both ground-truth APIs. All other methods miss at least one, with SEHGN notably failing to recommend ebay in the top-5. Neither WARBERT(R) nor WARBERT(M) alone successfully retrieves google-desktop, highlighting the necessity of the hierarchical architecture for handling specialized API combinations.

Overall, WARBERT consistently achieves the highest number of correct hits across diverse mashup categories including weather, music, web analytics, e-commerce, confirming that its hierarchical filter-then-match architecture and attention-based comparison mechanism enable effective semantic alignment between mashup requirements and API functionalities.

\subsection{Limitations}
Although WARBERT achieves strong overall performance, it also
faces several systematic limitations.

\textbf{Filtering-stage bottleneck.} The hierarchical architecture
introduces an error propagation risk: once WARBERT(R) excludes a relevant
API from the top-$H$ candidate set, WARBERT(M) cannot recover it. Among
the complete failure cases (zero correct API in top-10), 81.0\% originate
from this filtering-stage miss.

\textbf{Long-tail API coverage.} For frequently invoked APIs (training
frequency $>$ 30), WARBERT achieves Recall@5 of 89.7\%, whereas for
infrequently invoked APIs (frequency $\leq$ 5), Recall@5 drops to 31.6\%.
WARBERT(R) tends to default to popular but incorrect alternatives even
when the mashup description explicitly references the target API functions, for
example, predicting facebook for a mashup mentioning ``wookmark api show popular image.''

\textbf{Pairwise matching without set awareness.} WARBERT(M) evaluates
each mashup--API pair in isolation without modeling inter-API
complementarity. For example, tubegraph integrates youtube (video),
google-app-engine (infrastructure), and google-chart (visualization).
WARBERT(M) assigns a high score to video APIs including youtube, vimeo and dailymotion but pushes google-app-engine
to rank 7 and google-chart to rank 12, as the pairwise formulation cannot
recognize that the video role is already covered. 

These observations suggest that incorporating API co-invocation patterns
or structural metadata could further improve recommendation accuracy for
such challenging cases.

\subsection{Summary}

WARBERT achieves the best performance in all metrics. WARBERT shows its superiority over the baseline models and also demonstrates the effect of the hierarchical architecture by showing improvements over its component WARBERT(R) and WARBERT(M). We verify the effectiveness of the components of the WARBERT by ablation study and it shows the effect of the dual-component feature fusion and attention comparison strategy. We further investigate the impact of the candidate number $H$ and the weighting factor $\lambda$ on the inference time and performance of the model. The case study
illustrates WARBERT's recommendation behavior on representative mashups,
and the limitations analysis identifies systematic failure patterns.

\section{Conclusions and future work}
\label{sec:conclu}
The WARBERT framework, with its architecture of WARBERT(R) as an efficient filter and WARBERT(M) for precise matching, has demonstrated exceptional effectiveness in the Web API recommendation task. It consistently outperforms existing methods by achieving superior metrics such as Precision, Recall, NDCG, and MAP, with enhancements exceeding 9. 6\% to 11.7\% compared to MTFM. The proposed attention comparison mechanism and dual-component feature fusion enable a more profound understanding of API and mashup descriptions, which contribute to the framework's outstanding performance. Experimental results highlight the framework's advantage when combining both components, as opposed to using them individually. The ablation study illustrates the importance of the dual-component feature fusion and attention comparison and insights into the impact of the weighting factor $\lambda$ and the candidate number $H$. 

Looking ahead, WARBERT holds substantial potential for future enhancements, including enriching information sources beyond mashup and API descriptions, and expanding the capacity of the underlying language model.   As future work, we will replace the fixed fusion weight 
$\lambda$ with an input dependent, learnable gating mechanism to adaptively balance WARBERT(R) and WARBERT(M). We will also evaluate WARBERT on larger, multilingual datasets when such datasets become available to verify its generalizability.

\section*{Acknowledgment}
This work is by the Major Research Plan of the National Natural Science Foundation of China (Grant No.92582111), and the National Key Research and Development Program of China under Grant No.2022YFB4502001. The computation is completed in the HPC Platform of Huazhong University of Science and Technology.


\bibliographystyle{IEEEtran}
\bibliography{reference}

\begin{IEEEbiography}[{\includegraphics[width=1in,height=1.25in,clip,keepaspectratio]{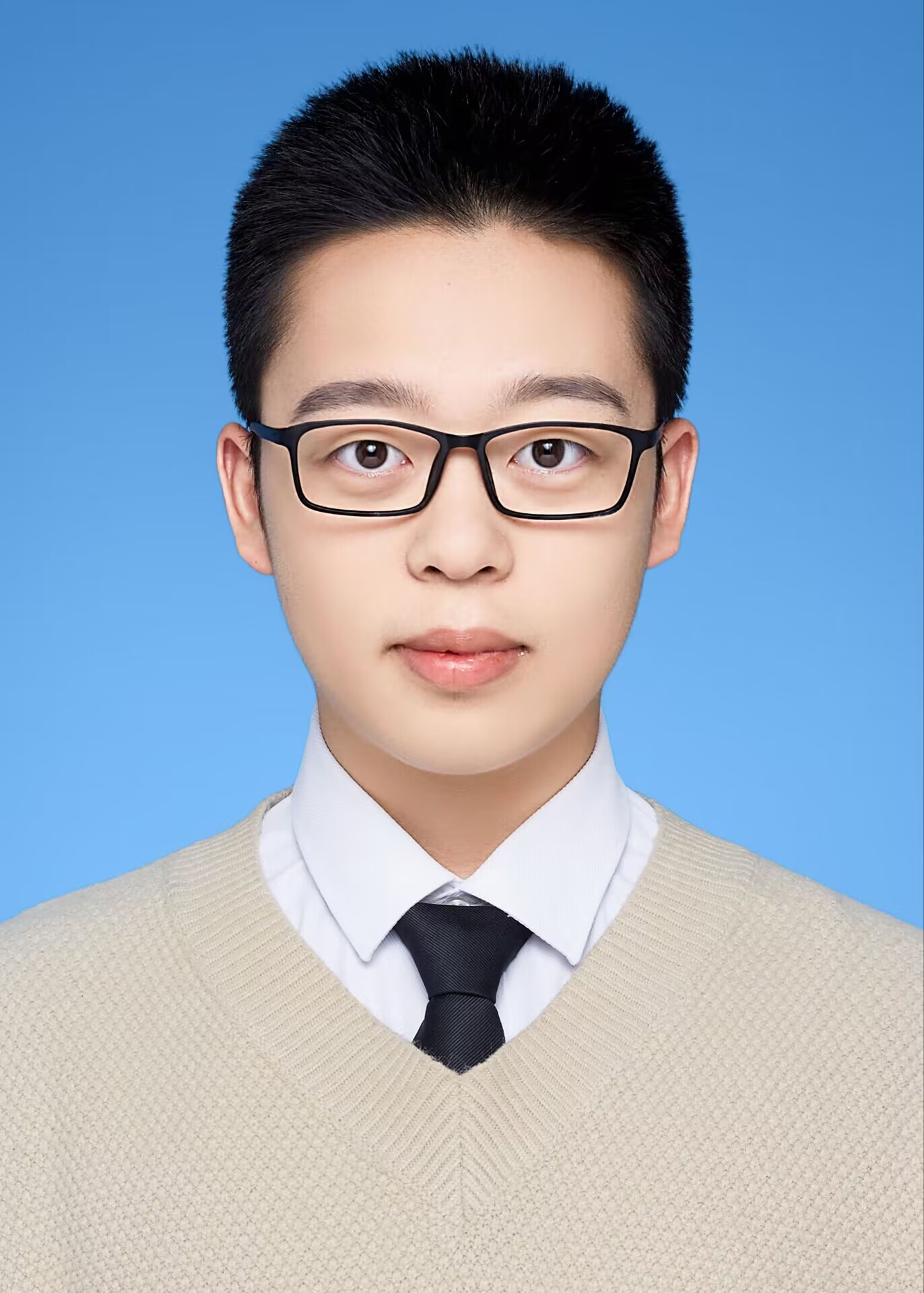}}]
{Zishuo Xu}  is working towards a B.S. degree in software engineering at the Huazhong University of Science and Technology (HUST), Wuhan, China. His research interests are machine learning and data mining.
\end{IEEEbiography}

\begin{IEEEbiography}
[{\includegraphics[width=1in,height=1.25in,clip,keepaspectratio]{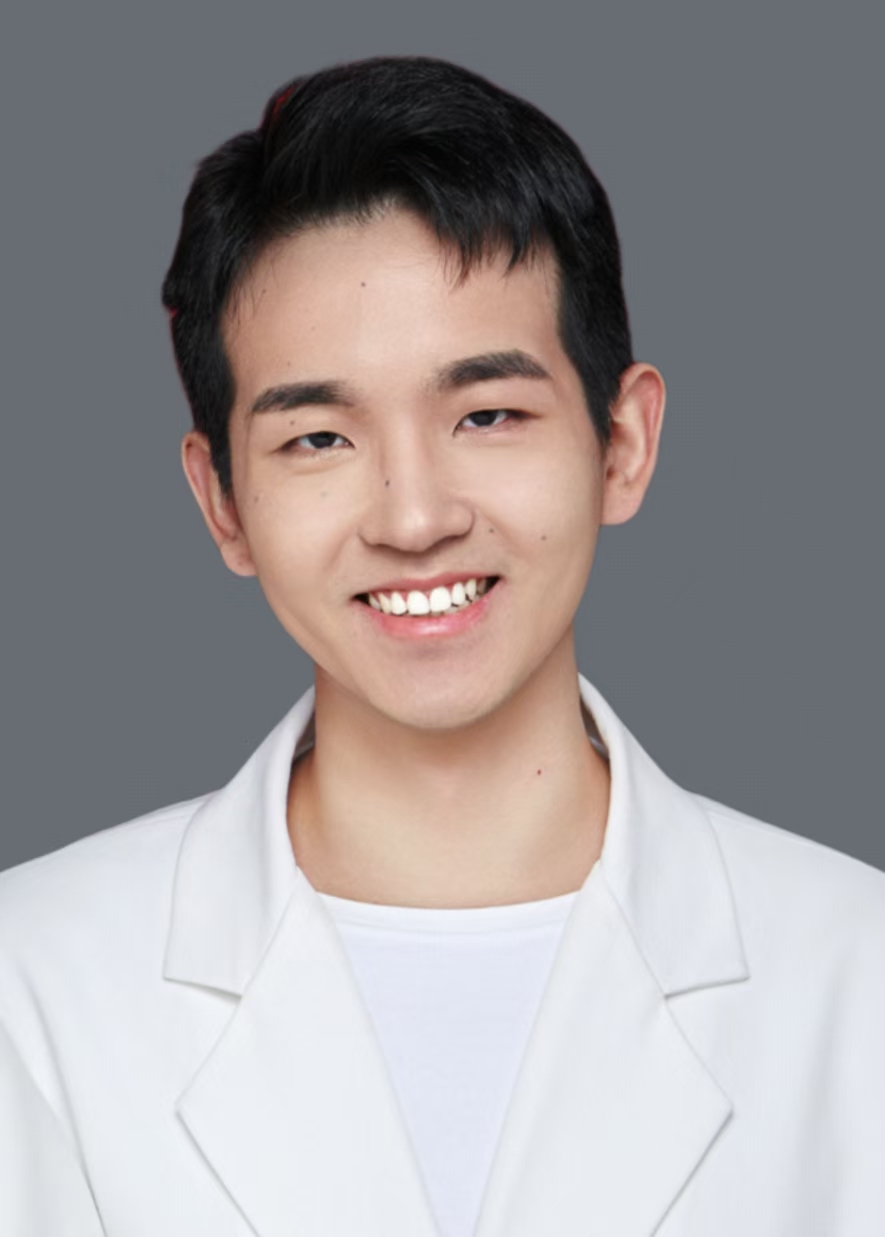}}]
{Yuhong Gu} received the B.S. degree from Northwestern Polytechnical University, Xi'an, China, in 2019, and the M.S. degree from Huazhong University of Science and Technology(HUST), Wuhan, China, in 2022. His main research interest is entity linking. He currently works for SmartX.

\end{IEEEbiography}

\begin{IEEEbiography}[{\includegraphics[width=1in,height=1.25in,clip,keepaspectratio]{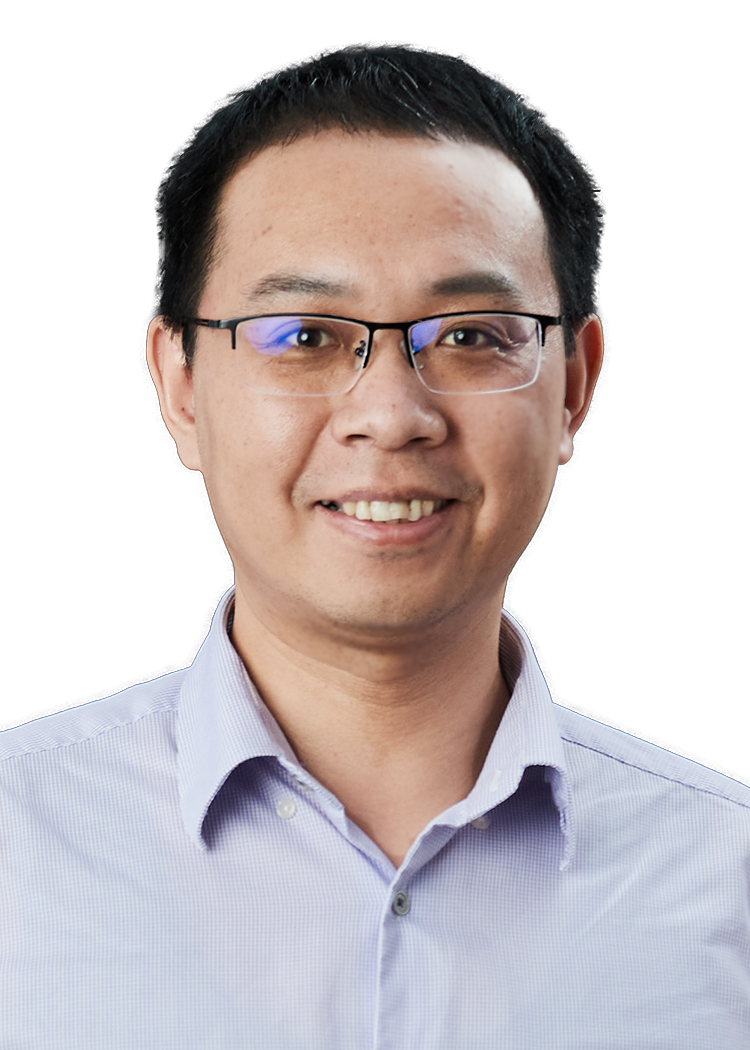}}]{Dezhong Yao}
	(Member, IEEE) received the PhD degree in computer science from the Huazhong University of Science and Technology (HUST), Wuhan, China, in 2016. He was a research fellow at Nanyang Technological University, Singapore, between 2016 and 2019. He is currently an Associate Professor with the School of Computer Science and Technology, HUST. His research interests include edge intelligence, cloud native, distributed computing, and federated learning. He has published more than 40 papers in top-tier conferences and journals, such as SIGMOD, WWW, NeurIPS, AAAI, IJCAI, FSE, ICPP, ICDM, TC, TPDS, TACO, TKDD, TCC and TIST.
\end{IEEEbiography}

\end{document}